\documentclass[preprint,12pt,nofootinbib]{revtex4-1}
\usepackage[paper=letterpaper,margin=1.5in]{geometry}
\usepackage[active]{srcltx} %SRC Specials for DVI search
\usepackage{amssymb,amsmath,amsfonts}
\usepackage{graphicx}
\usepackage{subfigure}
\usepackage{xcolor}

\newcommand{\be}{\begin{equation}}
\newcommand{\ee}{\end{equation}}
\renewcommand{\Re}{{\rm Re\,}}

\begin{document}

\begin{flushright}
{\normalsize
SLAC-PUB-15951\\
LCLS-II-TN-14-06\\
May 2014}
\end{flushright}

\vspace{.8cm}

\title{Roughness Tolerance Studies for the Undulator\\ Beam Pipe Chamber of LCLS-II}

\author{K. Bane and G. Stupakov}
%\date{\today}

\maketitle

\section*{Introduction}

When a bunch passes through the undulator of LCLS-II, the wakefields of the vacuum chamber will result in an added energy variation along the bunch, one that can negatively impact the FEL performance. The wakefield of the vacuum chamber is primarily due to the resistance of the walls and the roughness of the surface. To minimize the impact of the wakes, one would like a wall surface smooth enough so that the roughness component of the wake is a small fraction of the total wake. In LCLS-I, with an undulator vacuum chamber of the same material (aluminum) and roughly the same aperture as proposed for LCLS-II, the wall roughness tolerance specified as an rms slope of the surface of $(y')_{rms}=10$--15~mr was difficult to achieve~\cite{requirements}. The goal of this study is to understand the consequences to LCLS-II of loosening the roughness specification, say by a factor of 2 to 30~mr.

\begin{figure}[htb]
\centering
\includegraphics*[height=73mm]{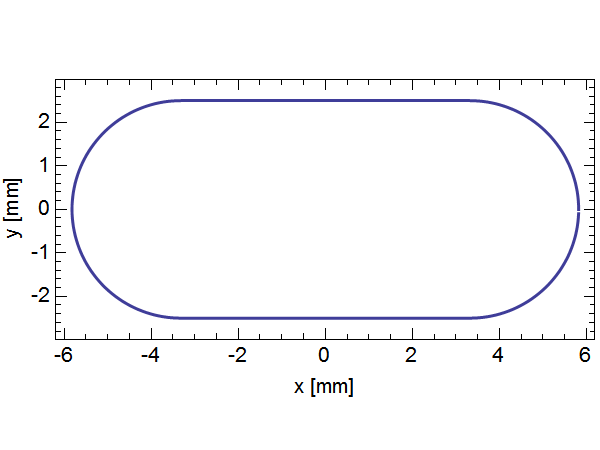}
\caption{Cross-section geometry of the undulator vacuum chamber.}
\label{geometry_fi}
\end{figure}

The vacuum chamber within the undulator of LCLS-II will be primarily extruded aluminum with a racetrack cross-section, as shown in Fig.~\ref{geometry_fi} (in addition, there are short breaks at the quads that will have a different shape and have a larger aperture). The full aperture is 5~mm by 12~mm, vertical by horizontal. From an impedance point of view, with the beam on axis, the effect is essentially the same as for the case of flat geometry, {\it i.e.} for a chamber consisting of two parallel plates with a vertical separation of 5~mm.

In this note we begin with the round approximation, {\it i.e.} we consider an aluminum pipe of radius $a=2.5$~mm. We calculate the total wake effect of resistive wall plus a model of roughness. The roughness model we use consists of small, shallow, sinusoidal corrugations~\cite{Stupakov2000}. We choose this model because measurements of samples of polished aluminum, similar to that to be used in the undulator chamber, find that the typical measured roughness is shallow~\cite{Stupakov99}.  Note that this model does not include a so-called ``synchronous mode" wake~\cite{NovokhatskiMosnier,BaneNovokhatski}. Such a mode appears in the case of small deep corrugations which are not expected for the LCLS II undulator.

The calculation of the short range wake of a resistive pipe has been done before~\cite{BaneSands}, as has the case of a pipe with small, shallow corrugations~\cite{StupakovReiche}; in this report we properly combine the two effects.
%We begin with a round model of the chamber because it is simpler, and because in the round case we can easily verify our analytical calculations with the 2D time-domain wakefield program ECHO \cite{ECHO}. 
Besides the analytical calculation, we present a simple way of estimating the relative contribution of the resistive wall and roughness components on the induced energy variation in the LCLS-II bunch. We next verify our analytical calculations with the 2D time-domain wakefield program ECHO~\cite{ECHO}.  Finally, we perform the corresponding analytical wake calculations for a vacuum chamber of flat geometry representing the LCLS-II vacuum chamber for different amounts of roughness.

Selected beam and machine properties in the undulator region of LCLS-II that are used in our calculations are given in Table~I.

\begin{table}[hbt]
   \centering
   \caption{Selected beam and machine properties in the undulator region of LCLS-II that are used in our calculations. The bunch charge given here is the maximum bunch charge to be used in LCLS-II. The longitudinal bunch distribution is approximately uniform.}
   \begin{tabular}{||l|c|c||}\hline 
        {Parameter name} & {Value}  &  Unit\\ \hline\hline 
       Charge per bunch, $Q$       &300  &pC  \\
      Beam current, $I$       &1  &kA  \\
       Rms bunch length, $\sigma_z$       &25  &$\mu$m \\
%      Repetition rate, $f_{rep}$       &1  &MHz \\
      Beam energy, $E$       &4  &GeV \\
      Vacuum chamber half aperture, $a$       &2.5  &mm \\      
      Vacuum chamber length, $L$       &130  &m \\
         \hline \hline 
   \end{tabular}
   \label{table1_tab}
\end{table}

\section*{Round Vacuum Chamber}

Consider first a round chamber of radius $a$, with wall resistance and small (in amplitude), shallow sinusoidal corrugations that represent the wall roughness. While in some cases the beam impedance can be calculated as a sum of the impedances due to resistance and that due to wall roughness, in general case such summation of impedances is not correct. A more general approach is based on the concept of surface impedance~\cite{stupakov1999} defined as the ratio of the longitudinal electric field and the azimuthal magnetic field at the wall, $\zeta=-(E_z/Z_0H_\phi)|_{r=a}$. Denoting $\zeta_{rw}(k)$ the wall resistive surface impedance and $\zeta_{ro}(k)$ the surface impedance due to roughness we can write the beam impedance $Z(k)$ as
\begin{equation}
Z(k)=\frac{Z_0}{2\pi a}\left({\frac{1}{\zeta_{rw}(k)+\zeta_{ro}(k)}-\frac{ik a}{2}}\right)^{-1}\ ,\label{Z_eq}
\end{equation}
with wave number $k=\omega/c$, where $\omega$ is frequency and $c$ is speed of light; and with $Z_0=377$~$\Omega$. The resistive wall surface impedance $\zeta_{rw}(k)$, is given by~\cite{AChao}
\begin{equation}
\zeta_{rw}(k)={(1-i)}\sqrt{\frac{k(1-ikc\tau_c)}{2Z_0\sigma_c}}\ ,
\end{equation}
with $\sigma_c$ the dc conductivity and $\tau_c$ the relaxation time of the metallic walls. The roughness surface impedance term is given by \cite{StupakovReiche}
\begin{equation}
\zeta_{ro}(k)=\frac{1}{4}kh^2\kappa^{3/2}\left(\frac{\sqrt{2k+\kappa}-i\sqrt{2k-\kappa}}{\sqrt{4k^2-\kappa^2}}\right)\ ;
\end{equation}
here the wall profile radius $r$ is assumed to vary sinusoidally with longitudinal position $z$: $r=h\cos \kappa z$. For the model to be valid we require the oscillations to be small and shallow, {\it i.e.} $\kappa a\ll1$ and $h\kappa\ll1$. Note that Eq.~\ref{Z_eq} implies that at low frequencies the two contributions to the impedance simply add:
\begin{equation}
Z(k)\approx\frac{Z_0}{2\pi a}\left[{\zeta_{rw}(k)+\zeta_{ro}(k)}\right]\quad\quad\quad\quad\quad{(ka\ll1)}\ ;
\end{equation}
however, as was pointed out above, in general this is not true. 
Once the impedance is known, then the wake is obtained by the inverse Fourier transform:
\begin{equation}
W_\delta(s)=\frac{c}{2\pi}\int_{-\infty}^\infty Z(k)e^{-iks}dk\ ,\label{IFT_eq}
\end{equation}
with $s$ the distance the test particle is behind the driving particle.
%Because of the smallness of the wall resistivity and the corrugation size,
%for efficient numerical calculation of Eq.~\ref{IFT_eq}, the integration variable $k$ needs to be properly scaled {\color{red} [this is not clear]}.
Note that in Ref.~\cite{StupakovReiche}
% for the case of a perfectly conducting pipe with roughness corrugations, 
further practical considerations for such a calculation as a contour integral are discussed.
% and an efficient calculation method is presented. 

For the LCLS-II undulator vacuum chamber the dominant effect is expected to be the resistive wall wake, with the roughness corrugations contributing to a lesser degree. 
The strength of the resistive wall wake for a short bunch depends on the characteristic distance
\begin{equation}
s_0=\left(\frac{2a^2}{Z_0\sigma_c}\right)^{1/3}\ ,\label{s0_eq}
\end{equation}
which represents a location near the first zero crossing of the point charge wake. 
For aluminum the conductivity $\sigma_c=3.5\times10^7$~$\Omega^{-1}$m$^{-1}$ and relaxation time $\tau_c=8$~fs; with $a=2.5$~mm, $s_0=9.8$~$\mu$m. For very short bunches it is $s_0$ rather than $\sigma_c^{-1/2}$ that gives the scale of the strength of the wake in a bunch. 
For the roughness model, the long range wake is given by \cite{Stupakov2000}
\begin{equation}
W_\delta(s)=-\frac{Z_0c}{16\pi^{3/2}a}\frac{h^2\kappa^{3/2}}{s^{3/2}}=-\frac{c}{4\pi^{3/2}}\sqrt{\frac{Z_0}{(\sigma_c)_{ro}}}\frac{1}{s^{3/2}}\ ,\label{roughness_wake_eq}
\end{equation}
with the overall minus sign in the expression indicating that the test particle gains energy from the leading particle\footnote{We define the sign of the wake so that the positive wake corresponds to the energy gain.}. This is the same $s$ dependence as for the long range resistive wall wake, and in the second expression on the right we write the wake in terms of an equivalent roughness conductivity
\begin{equation}
(\sigma_c)_{ro}=\frac{16}{Z_0h^4\kappa^3}\ .\label{sigma_ro_eq}
\end{equation}
Inserting this conductivity into Eq.~\ref{s0_eq}, one obtains an effective roughness distance $(s_0)_{ro}$. Choosing $\lambda_{ro}=2\pi/\kappa=300$~$\mu$m, $(y')_{rms}=h\kappa/\sqrt{2}=30$~mr, we find that $(\sigma_c)_{ro}=2.9\times10^8$~$\Omega^{-1}$m$^{-1}$ and $(s_0)_{ro}=4.9$~$\mu$m. We see that the characteristic distance for this level of wall roughness is about half that of the wall resistance.

\begin{figure}[htbp]
\centering
\includegraphics*[height=73mm]{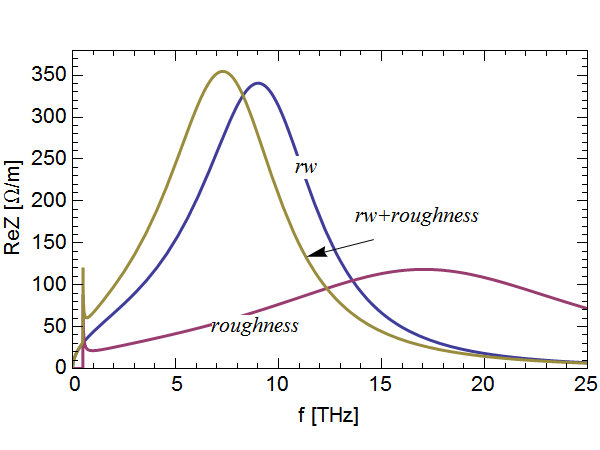}
\includegraphics*[height=73mm]{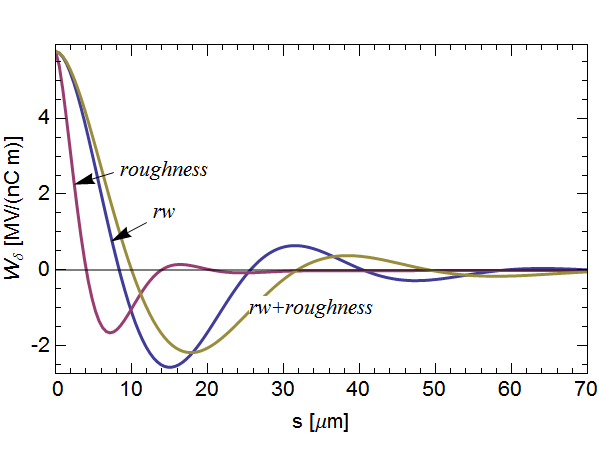}
\caption{For round geometry: $\Re\! Z(f)$ (top) and point charge wake $W_\delta(s)$ (bottom) for the case of a pipe that has resistance (blue), roughness (red), and both resistance and roughness (yellow). The resistive wall calculation includes ac conductivity for aluminum; the roughness model assumes $(y')_{rms}=30$~mr and wall oscillation wavelength $\lambda_{ro}=300$~$\mu$m.}
\label{round_wake_fi}
\end{figure}

We numerically performed the integral of Eq.~\ref{IFT_eq}, considering the effects of the resistivity of aluminum, and the wall roughness with $(y')_{rms}=30$~mr and oscillation wavelength $\lambda_{ro}=300$~$\mu$m. In Fig.~\ref{round_wake_fi} we present $\Re\! Z(f)$ (top; $f$ is frequency) and the point charge wake $W_\delta(s)$ (bottom) for the case of a pipe that has wall  resistance (blue), roughness (red), and both resistance and roughness (yellow). We see that the total effect is dominated by the resistive wall wake, and it is not simply given by the sum of the two individual wakes. We further note that $W_\delta(0^+)=Z_0c/\pi a^2=5.8$~MV/(nC m). The first zero-crossing of the wakes is near $s_0=9.8$~$\mu$m, $(s_0)_{ro}=4.9$~$\mu$m, and $(s_0)_{tot}=12$~$\mu$m, respectively, where the combined effect has been approximated by
\begin{equation}
(s_0)_{tot}=\left(\frac{1}{\sigma_c^{1/2}}+\frac{1}{(\sigma_c)_{ro}^{1/2}}\right)^{-2}\ .\label{s0_tot_eq}
\end{equation}

\begin{figure}[htb]
\centering
\includegraphics*[height=73mm]{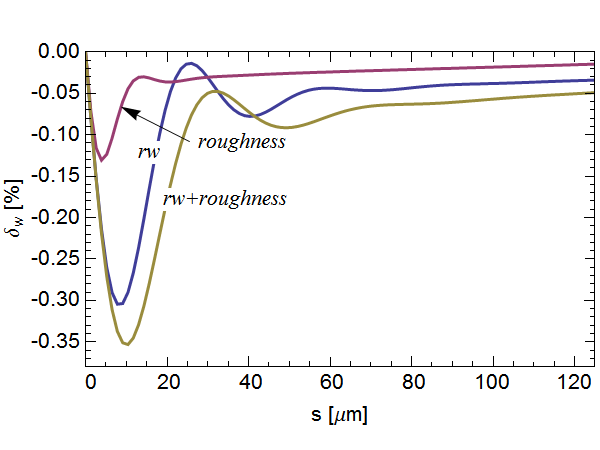}
\caption{For round geometry: relative induced voltage $\delta_w(s)$ along a bunch with a uniform distribution,  for the case of a pipe that has resistance (blue), roughness (red), and both resistance and roughness (yellow). The resistive wall calculation includes ac conductivity for aluminum; the roughness model assumes $(y')_{rms}=30$~mr and wall oscillation wavelength $\lambda_{ro}=300$~$\mu$m. The beam has a uniform distribution and with its head located at $s=0$. The beam peak current $I=1$~kA,  and the beam reaches to 30 (90)~$\mu$m for the $Q=100$ (300)~pC case. The energy $E=4$~GeV, and the length of pipe $L=130$~m.}
\label{round_induced_fi}
\end{figure}

In the undulator region of LCLS-II the longitudinal bunch distribution is roughly uniform, with peak current $I=1$~kA; the nominal bunch charge is $Q=100$~pC, with a maximum of $Q=300$~pC possible. The bunch wake is given by the convolution
\begin{equation}
W_\lambda(s)=-\int_0^\infty W_\delta(s')\lambda(s-s')\,ds'\ ,\label{convolve_eq}
\end{equation}
with $\lambda(s)$ the longitudinal bunch distribution, and a negative value for $W_\lambda(s)$ indicates energy loss. For a uniform bunch distribution with peak current $I$ the relative wake induced energy variation at the end of the undulator is given by
\begin{equation}
\delta_w(s)=-\frac{eIL}{cE}\int_0^s W_\delta(s')\,ds'\ ,
\end{equation}
with $L$ the length of the undulator pipe and $E$ the beam energy. In Fig.~\ref{round_induced_fi} we plot the relative induced voltage in a uniform bunch for the three cases of Fig.~\ref{round_wake_fi}. We see that for both the 100~pC bunch (total length of $\ell=2\sqrt{3}\sigma_z=30$~$\mu$m) and the 100~pC bunch ($\ell=90$~$\mu$m) the total energy variation induced within the bunch is $\Delta\delta_w=0.36$\% for resistance plus roughness, {\it vs.} 0.30\% for resistance without roughness; the roughness adds a 20\% effect. Since the wake drops nearly linearly to zero near the effective $s_0$, we can estimate these numbers with the formula
\begin{equation}
\Delta\delta_w=\frac{Z_0\bar s_0}{2\pi a^2}\frac{eIL}{E}\ ,\label{Deltaw_est_eq}
\end{equation}
where $\bar s_0=(s_0)_{tot}$ in the former case, or $\bar s_0=s_0$ in the latter one;
which gives $\Delta\delta_w=0.37$\% and 0.31\% for, respectively, the case of roughness plus resistance, and the case of resistance alone---in good agreement to the more accurate calculations.

\section*{Numerical Tests}

\begin{figure}[htb]
\centering
\includegraphics*[height=73mm]{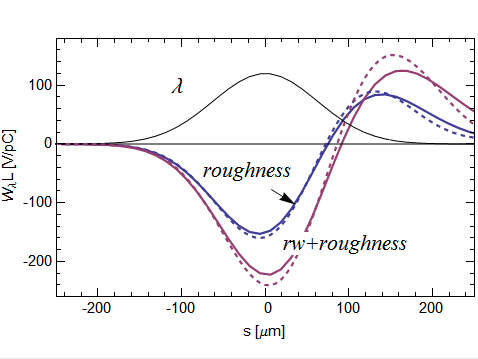}
\caption{Comparisons of bunch wake as obtained by ECHO (solid curves) and by the analytical calculations (dashes) for two test examples: (1)~a lossless, corrugated pipe (blue), and (2)~a lossy, corrugated pipe (red). The bunch shape $\lambda$ is also shown, with the head to the left.}
\label{echo_tests_fi}
\end{figure}

We next present test calculations for the round geometry with the finite difference wakefield code ECHO. This code can calculate the effects of both geometric and resistive wall (dc only) wakes (provided that the skin depth is small compared to the size of the wall perturbations). However, a sinusoidal wall oscillation as small as {\it e.g.} $(y')_{rms}=30$~mr on an $a=2.5$~mm pipe is difficult to simulate, so we artificially enlarged the oscillations and reduced the wall conductivity. We consider two cases: (1)~roughness alone and (2)~roughness plus wall resistance. Parameters are: $a=2.5$~mm, $\lambda_{ro}=2.5$~mm, $h=60$~$\mu$m, pipe length $L=25$~cm, wall conductivity $\sigma_c=6\times10^5$~$\Omega^{-1}$m$^{-1}$; so $s_0=37$~$\mu$m and $(y')_{rms}=110$~mr. The bunch is Gaussian with rms length $\sigma_z=60$~$\mu$m and the skin depth $\delta_s=0.8$~$\mu$m. The mesh size was taken to be 12~$\mu$m. For analytical comparison to the ECHO results we inserted Eq.~\ref{Z_eq} into Eq.~\ref{IFT_eq} to find the point charge wake. This function was convolved according to Eq.~\ref{convolve_eq} to obtain the bunch wake. The results are shown in Fig.~\ref{echo_tests_fi}. The ECHO results are given by the solid curves, and the analytic results by dashes. We see good agreement.

\section*{Flat Vacuum Chamber}

\begin{figure}[htbp]
\centering
\includegraphics*[height=73mm]{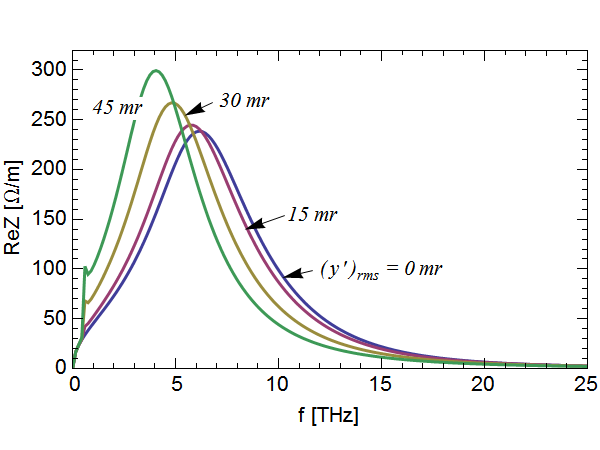}
\includegraphics*[height=73mm]{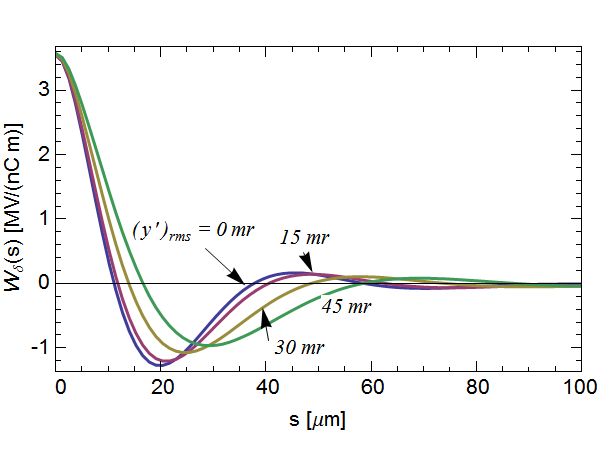}
\caption{For flat geometry: $ReZ(f)$ (top) and $W_\delta(s)$ (bottom) for the ac resistance model of aluminum plus the effects of the roughness model, for the cases $(y')_{rms}=0$, 15, 30, 45~mr. Here the wall oscillation wavelength $\lambda_{ro}=300$~$\mu$m.}
\label{flat_imp_fi}
\end{figure}

Henke and Napoly give the impedance of a resistive wall in flat geometry in Ref.~\cite{Henke}. With a slight modification we include the effects of both the wall resistance and roughness:
\begin{equation}
Z(k)=\frac{Z_0}{2\pi a}\int_{-\infty}^\infty dq\,{\rm sech}\, q\left({\frac{\cosh q}{\zeta_{rw}(k)+\zeta_{ro}(k)}-\frac{ik a}{q}\sinh q}\right)^{-1}\ .\label{Zflat_eq}
\end{equation}
We have repeated the previous calculations for flat geometry, for cases of aluminum with ac conductivity and roughness with $(y')_{rms}$ of: (1)~0~mr, (2)~15~mr, (3)~30~mr, and (4)~45~mr ($\lambda_{ro}=300$~$\mu$m in all cases). The resulting impedances are shown in Fig.~\ref{flat_imp_fi} (top), the point charge wakes in Fig.~\ref{flat_imp_fi} (bottom). We see that, compared to the round case, $W(0^+)$ is reduced by the factor $\pi^2/16$ and the first zero crossing of the wake is increased slightly.
Thus Eqs.~\ref{sigma_ro_eq}, \ref{s0_tot_eq}, and \ref{Deltaw_est_eq}---with the last one multiplied by $\pi^2/16$---can still be used to estimate the relative impact of the roughness and the wall resistance.
%We see that, as in the round case, adding roughness reduces the central frequency and increases the amplitude of the broad resonance in $ReZ(f)$, and it increases the first zero crossing location in $W_\delta(s)$. 

\begin{figure}[htb]
\centering
\includegraphics*[height=73mm]{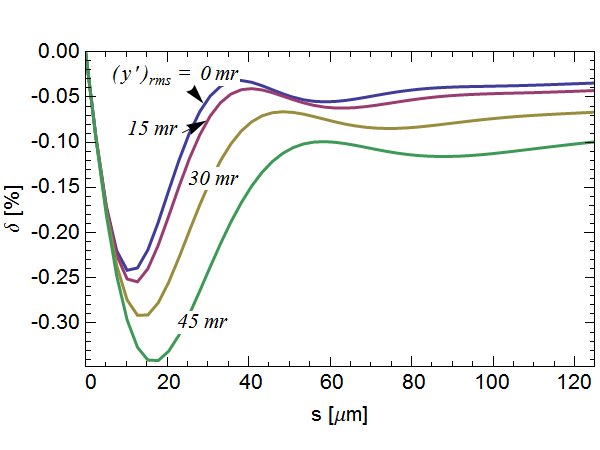}
\caption{For flat geometry: relative induced voltage $\delta_w(s)$ along the bunch, for the cases wall resistance plus roughness, with $(y')_{rms}=0$, 15, 30, 45~mr (the wall oscillation wavelength $\lambda_{ro}=300$~$\mu$m). The beam has a uniform distribution with its head located at $s=0$. The peak current $I=1$~kA,  and the beam reaches to 30 (90)~$\mu$m for the $Q=100$ (300)~pC case. The energy $E=4$~GeV, and the length of pipe is $L=130$~m.}
\label{flat_induced_fi}
\end{figure}

 In Fig.~\ref{flat_induced_fi} we plot the relative induced energy variation for a uniform bunch distribution. Here the peak current $ I=1$~kA; the bunch head is located at $s=0$, with the entire bunch extent reaching to 90~$\mu$m (for the $Q=300$~pC case), and to 30~$\mu$m (for the nominal $Q=100$~pC case). The length of pipe is assumed to be $L=130$~m, and the beam energy $E=4$~GeV.
The total induced relative energy variation for a resistive pipe with no roughness (for both the 100~pC and 300~pC cases) is $\Delta\delta_w=0.25$\%. Adding roughness increases this value by 5\%, 19\%, 38\%, when $(y')_{rms}=15$, 30, 45~mr, respectively.

The roughness effect depends on $(y')_{rms}$ and also on $\lambda_{ro}$, though the latter dependence is expected to be much weaker. Repeating the calculation for wall resistance plus roughness with $(y')_{rms}=30$~mr, but taking $\lambda_{ro}=900$~$\mu$m we find that the roughness increases $\Delta\delta_w$ by 27.5\% compared to the effect of wall resistance alone. This confirms that the dependence of $\Delta\delta_w$ on $\lambda_{ro}$ is weak.

\begin{figure}[htbp]
\centering
\includegraphics*[height=73mm]{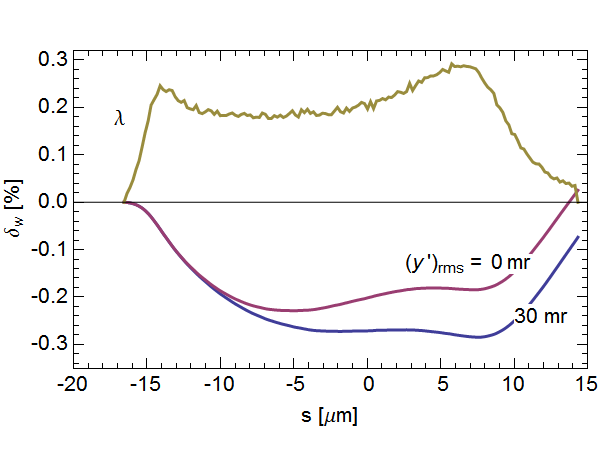}
\includegraphics*[height=73mm]{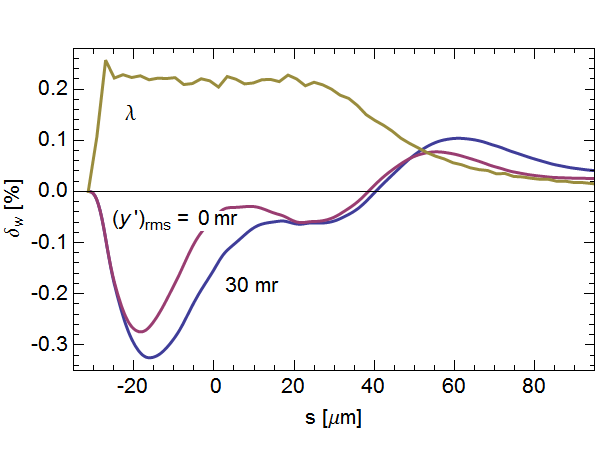}
\caption{For flat geometry, relative induced voltage $\delta_w$ for numerically obtained bunch shapes, for $Q=100$~pC (top) and $Q=300$~pC (bottom). The curves represent the effect of ac resistivity in aluminum (red)  and that of resistance plus roughness with $(y')_{rms}=30$~mr (blue). Here the wall oscillation wavelength $\lambda_{ro}=300$~$\mu$m. The bunch shapes are given in yellow, with the head to the left.}
\label{flat_induced_real_fi}
\end{figure}

The LCLS-II bunch distribution in the undulator is not exactly uniform with peak current $I=1$~kA (see Fig.~\ref{flat_induced_real_fi}, the yellow curves). The current, numerically obtained 100~pC distribution has slight horns at the head and tail of the bunch, with a slight current droop in the middle; the 300~pC distribution can be described as uniform in front with a long trailing tail. Repeating the induced energy spread calculations with these distributions, both with wall resistance alone and with resistance plus roughness with $(y')_{rms}=30$~mr ($\lambda_{ro}=300$~$\mu$m), we obtain $\delta_w(s)$ as given by the red and blue curves in Fig.~\ref{flat_induced_real_fi}.
It is interesting to note that, because the 300~pC bunch shape begins as a uniform distribution, $\delta_w(s)$ quickly drops and rises back to near zero, similar to the behavior in Fig.~\ref{flat_induced_fi}. For the 100~pC case, however, because of the horns and droop, $\delta_w(s)$---after reaching its minimum---remains flattened for most of the rest of the bunch.

For these calculations, we find that $\Delta\delta_w$ increases by 24\% (19\%) for $Q=100$~pC (300~pC),  comparing the case with roughness to the one without. These results are not far from the 19\% increase estimated above for the uniform distribution. Finally, for completeness, we calculate the wakefield-induced power loss in the undulator beam pipe: $P=\langle W_\lambda\rangle Q^2f_{rep}/L$, where $\langle\rangle$ indicates averaging over the bunch. We find that $P=2.1$ (1.0)~W/m for $Q=100$ (300)~pC, using the maximum planned repetition rate, $f_{rep}=300$ (100)~kHz.

\section*{Conclusions}

We have investigated the wake effect of the wall resistance and roughness of the undulator beam pipe on the LCLS-II beam. In particular we wanted to see if it is acceptable to loosen the roughness tolerance from an equivalent rms slope at the surface of $(y')_{rms}=15$~mr to 30~mr. According to the calculations presented here, such a loosening will result in the roughness contribution to the induced voltage to increase from 5\% to 20\%. The absolute scale of the total wake effect is a relative induced energy variation of $\sim0.3$\% (assuming a pipe length of 130~m and a beam energy of 4~GeV).

In this note, we have presented an analytical calculation of the wake in a round or flat chamber with wall resistance and shallow, sinusoidal corrugations. We have additionally shown that our analytical calculations of the short range wake in such a chamber is in good agreement with results of the time domain, finite difference program ECHO. Finally, we have presented a simple model for estimating the extra effect of wall roughness on the wake of the beam in the LCLS-II undulator chamber.

\section*{Acknowledgement}
The authors thank L. Wang for providing the bunch shapes shown in Fig.~\ref{flat_induced_real_fi}. Work
supported by Department of Energy contract DE--AC02--76SF00515.

\end{document}